\documentclass[nofootinbib,reprint,amsmath,amssymb,aps]{revtex4-1}
\usepackage[utf8,latin1]{inputenc}
\usepackage{graphicx}
\usepackage[dvipsnames]{xcolor}
\usepackage{dcolumn}

\usepackage{mathabx}
\usepackage{mathrsfs}  
\usepackage{tensor}
\usepackage[normalem]{ulem}
\usepackage{cancel}
\usepackage{bm}

\usepackage{hyperref}
\usepackage{stackengine,scalerel}
\hypersetup{colorlinks, linkcolor={red},citecolor={blue},urlcolor={blue}}

\def\nn{\nonumber} 

\newcommand{\dd}{{\rm d}}
\newcommand{\OO}{{\rm O}}
\newcommand{\OC}[1]{{\rm O}\left(c^{#1}\right)}

\begin{document}

\title[Analytical results for binary dynamics at the first post-Newtonian order in Einstein-Cartan theory with the Weyssenhoff fluid]{Analytical results for binary dynamics at the first post-Newtonian order\\ in Einstein-Cartan theory with the Weyssenhoff fluid}

\author{Vittorio De Falco$^{1,2}$}
\email{v.defalco@ssmeridionale.it}
\author{Emmanuele Battista$^{3}$} \email{emmanuele.battista@univie.ac.at}
\email[\\]{emmanuelebattista@gmail.com}

\affiliation{$^1$ Scuola Superiore Meridionale, Largo San Marcellino 10, 80138 Napoli, Italy,\\
$^2$ Istituto Nazionale di Fisica Nucleare, Sezione di Napoli, Complesso Universitario di Monte S. Angelo, Via Cintia Edificio 6, 80126 Napoli, Italy\\
$3$ Department of Physics, University of Vienna, Boltzmanngasse 5, A-1090 Vienna, Austria
}

\date{\today}

\begin{abstract}
The quantum spin effects inside matter can be modeled via the Weyssenhoff fluid, which permits to unearth a formal analogy between general relativity and Einstein-Cartan theory at the first post-Newtonian order. In this framework, we provide some analytical formulas pertaining to the dynamics of binary systems having the  spins aligned perpendicular to the orbital plane. We derive the expressions of the relative orbit and the coordinate time, which in turn allow to determine the gravitational waveform, and the energy and angular momentum fluxes. The potentialities of our results are presented in two  astrophysical applications, where we compute: ($i$) the quantum spin contributions to the energy flux and gravitational waveform during the inspiral phase; ($ii$) the macroscopic angular momentum of one of the bodies starting from the time-averaged energy flux and the knowledge of few timing parameters. 
\end{abstract}

\maketitle

\section{Introduction} 

When Einstein laid down the basis of general relativity (GR) in 1916, quantum mechanics had not yet been formalized. This means that quantum concepts like the spin have no geometrical counterpart in GR, which thus configures as a purely classical theory. The revolutionary ideas underlying this framework became soon a source of inspiration for several authors. In particular, in 1920 Cartan developed an extension of GR, now referred  to as \emph{Einstein-Cartan (EC) theory} \cite{Cartan1952}, where the most general metric-compatible affine connection was taken into account. This geometric formulation was then revisited by Kibble and Sciama in the sixties, who devised it within the gauge theory of the Poincar\'e group \cite{Kible1961, Sciama1962}. It was then realized that the torsion tensor $S^{\lambda}_{\ \mu \nu}$, i.e., the  antisymmetric part of the affine connection, is associated with the intrinsic quantum spin of matter \cite{Hehl1976_fundations,Hehl2012-review,Blagojevic2013}.

One of the chief differences between EC model and GR reside in their geometrical foundations, as the former is framed in the Riemann-Cartan environment, whereas the latter in the Riemannian arena \cite{Hehl1976_fundations}. This explains why EC pattern naturally fits the gauge paradigm, whereas if this is applied to Einstein gravity we end up with the teleparallel equivalent of GR (TEGR), i.e., the gauge theory of the translation group \cite{Aldrovandi-Pereira2013,Capozziello2022}.
Furthermore, EC field equations can be derived from a Palatini action over a Riemann-Cartan geometry, where the torsion is independent of the metric. In this way, the principle of least action yields ($\kappa:= 8 \pi G /c^4$) \cite{Hehl1976_fundations,Gasperini-DeSabbata} 
\begin{subequations} \label{EC-field-equations-1}
\begin{align}
& G^{\alpha \beta} = \kappa \mathbb{T}^{\alpha \beta},
\label{first-EC-field-equation}
\\ 
& S_{\mu \nu}{}^{\lambda} + \delta^\lambda_{\mu} S_{\nu\rho}{}^{\rho}-\delta^\lambda_{\nu} S_{\mu\rho}{}^{\rho} = \kappa \tau_{\mu \nu}{}^{\lambda},
\label{torsion&tau}
\end{align} 
\end{subequations}
where $G^{\alpha \beta}$ is the EC tensor. From the above equations it is clear that both mass and spin represent the source of gravitation, since $\mathbb{T}^{\alpha \beta}$ and $\tau_{\mu \nu}{}^{\lambda}$ denote the canonical stress-energy and spin angular momentum tensors, respectively. They are linked via the relation $\mathbb{T}^{\alpha \beta}=T^{\alpha \beta} + \left(\nabla_\gamma + 2 S_{\gamma \mu}{}^{\mu}\right) \left(\tau^{\alpha \beta \gamma}-\tau^{ \beta \gamma \alpha}+\tau^{\gamma \alpha \beta }\right)$, $T^{\alpha \beta}$ being the metric stress-energy tensor. A peculiar aspect of EC theory is that Eq. \eqref{torsion&tau} is an algebraic equation. As a consequence, \emph{torsion does not propagate and hence it is confined only to the region occupied by matter}. If one exploits Eq. \eqref{torsion&tau}, then EC field equations can be recast in the GR-like form \cite{Hehl1976_fundations} 
\begin{align}\label{EC-field-equations-2}
\hat{G}^{\alpha \beta}&=\kappa \Theta^{\alpha\beta}.
\end{align} 
Here, $\hat{G}^{\alpha \beta}$ is the (symmetric) Einstein tensor, which differs from $G^{\alpha \beta}$ due to the torsion contributions. Moreover, $\Theta^{\alpha\beta}:=T^{\alpha\beta}+ \kappa \mathcal{S}^{\alpha \beta}$ denotes the combined energy-momentum tensor, where $\mathcal{S}^{\alpha \beta}$ is what we dub torsional energy-momentum tensor, due to its (quadratic) dependence on $\tau_{\mu \nu}{}^{\lambda}$. It is worth noticing that Eq. \eqref{EC-field-equations-2} does not imply that EC theory is a trivial generalization of GR. In fact, the formulation \eqref{EC-field-equations-2} is useful from a practical point of view, but it does not entail we have abandoned the Riemann-Cartan arena. In other words, GR pertaining to bodies endowed with angular momentum and EC theory are not equivalent. Indeed, in GR macroscopic rotations engender the modification of the stress-energy tensor and not of the geometry, while in EC model both are affected by the presence of the quantum spin.
 
In EC theory, matter-field dynamics can be derived from the generalized conservation laws \cite{Hehl1974b,Hehl1976_fundations,Gasperini-DeSabbata}
\begin{subequations}
\label{generalized-conservation-laws}
\begin{align}  
\left(\nabla_\nu + 2 S_{\nu \alpha}{}^{\alpha}\right)\mathbb{T}_{\mu}{}^{\nu} &= 2 \mathbb{T}_{\lambda}{}^{\nu} S_{\mu \nu}{}^{\lambda} -  \tau_{\nu \rho}{}^{\sigma} R_{\mu \sigma}{}^{\nu \rho},
\label{conservation-law-energy-momentum}
\\
2\left(\nabla_\lambda + 2 S_{\lambda \alpha}{}^{\alpha}\right) \tau_{\mu \nu}{}^{\lambda} &= \mathbb{T}_{\mu \nu}-\mathbb{T}_{\nu \mu},
\label{conservation-law-angular-momentum}
\end{align}
\end{subequations}
where both the covariant derivative and the Riemann tensor include torsion contributions. The above relations  are a consequence of Eq. \eqref{EC-field-equations-1}. In particular, Eq. \eqref{conservation-law-energy-momentum} follows from the contracted Bianchi identity framed in Riemann-Cartan geometry, while Eq. \eqref{conservation-law-angular-momentum} originates from the antisymmetric part of $G^{\alpha \beta}$ \cite{Gasperini-DeSabbata}.   

The first physical features of the equations of motion can be figured out by considering a test particle. In EC framework, test-body  trajectories are neither geodesics nor autoparallel curves already at the pole-particle approximation, which yields in fact a set of Mathisson-Papapetrou-Dixon-like equations for the translational dynamics. These exhibit explicitly the contributions of the torsion tensor and contain a coupling term between the quantum spin of the object and the curvature of the spacetime \cite{Hehl1971,Gasperini-DeSabbata}. On the other hand, the standard Mathisson-Papapetrou-Dixon equations of GR refer to the macroscopic angular momentum of the body and are worked out in the pole-dipole approximation \cite{Papapetrou1951}. 

A more advanced scenario is represented by the dynamics of a self-gravitating system, which is fundamental in gravitational-wave (GW) theory. In our research program, we have studied the GW generation problem with the Blanchet-Damour formalism in EC theory by first considering a source shaped by the  Weyssenhoff fluid \cite{Paper1,Paper2}. This is characterized by the tensors \cite{Obukhov1987,Paper2}
\begin{subequations}
\begin{align}
\mathbb{T}^{\alpha \beta}&= p^\alpha u^\beta + \left(u^\alpha u^\beta/c^2 + g^{\alpha \beta} \right)P,
\\ 
\tau_{\alpha \beta}{}^{\gamma} &= s_{\alpha \beta} u^\gamma,
\label{spin-tensor-Weyssenhoff-fluid}
\end{align}    
\end{subequations}
where $p^\alpha$, $u^\alpha$, $P$, and $s_{\alpha \beta}$ are the four-momentum, four-velocity, pressure, and spin density tensor of the fluid, respectively. The conservation laws \eqref{generalized-conservation-laws} yield a generalized   Euler equation and precessional motion  showing significant deviations from the GR expectations. We have studied both the translational and the rotational fluid evolution via the post-Newtonian (PN) approximation scheme and by adopting the Frenkel condition $s_{\alpha\beta}u^\beta=0$ \cite{Paper2}. 
Then, we have dealt with compact binaries, which represent the main candidates for GWs in astrophysics. These can be formally described by applying the point-particle limit to the continuous fluid distribution. In this way, we have derived the equations of motion and the radiative multipole moments of a spinning PN two-body system \cite{Paper2,Paper3,Paper4}. 

\begin{figure*}[ht!]
    \centering
    \includegraphics[trim=0cm 4cm 0cm 4cm,scale=0.6]{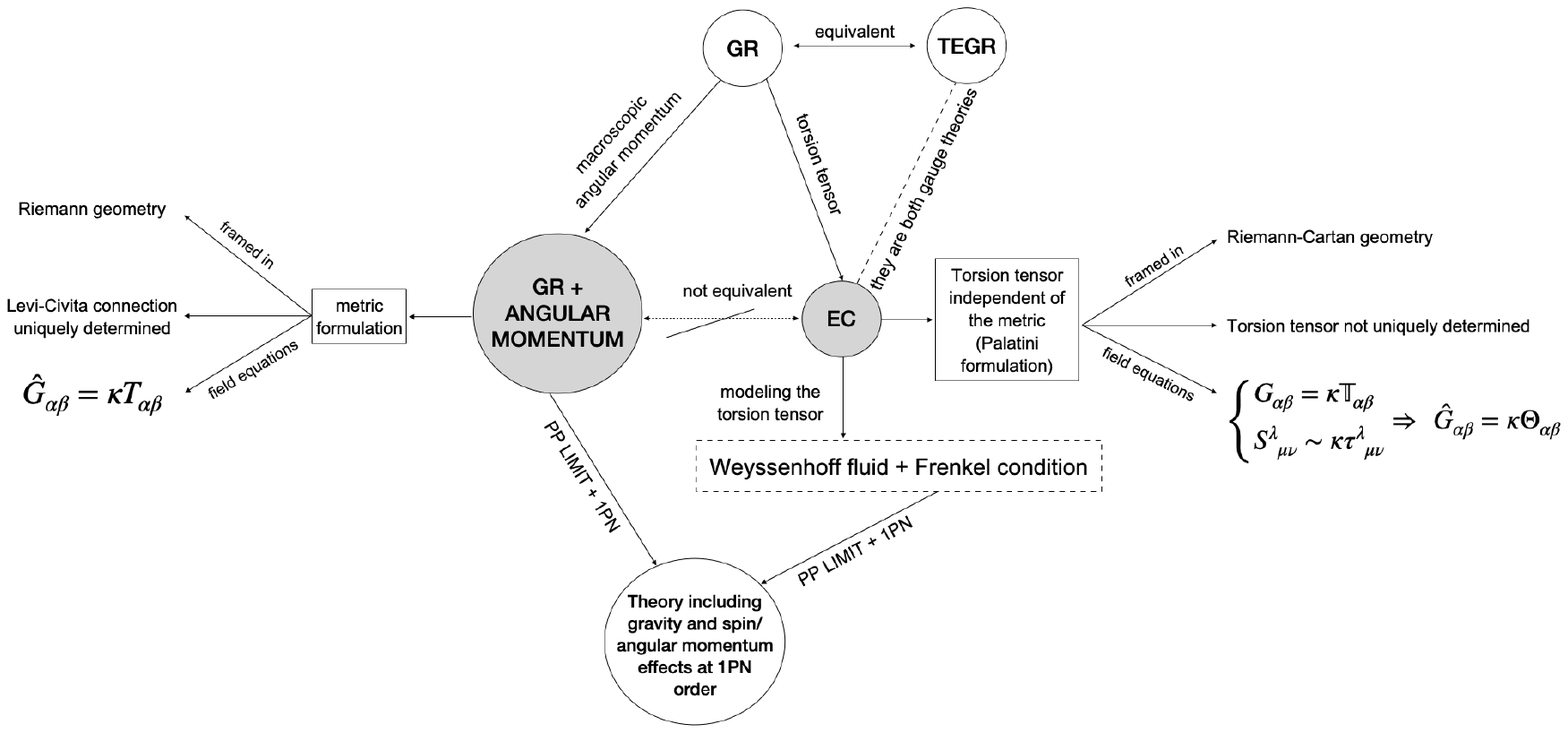}
    \caption{Scheme showing differences and formal analogies between the GR description of bodies endowed with macroscopic angular momentum and the EC treatment of objects having a quantum spin.  These two frameworks are not equivalent, because GR is a metric theory of gravity, whereas EC model follows a Palatini formulation.  Despite that, if we consider the Weyssenhoff fluid and the Frenkel condition to model the quantum spin effects in EC framework, we discover,  after having applied the point-particle limit (refereed to as \emph{PP limit}), that the two theories share some common facets at 1PN level.}
    \label{fig:schema2}
\end{figure*}

Our investigation regarding binary systems has revealed some remarkable novel results holding at the first post-Newtonian (1PN) level, which we summarise as follows:
\begin{enumerate}
\item both the translational and the rotational equations of motion resemble formally those of GR up to a normalization factor in the spin vector \cite{Paper3,Paper4}; 
\item the effacing principle is valid. This conclusion has been achieved after a careful investigation of the inner-structure-dependent integrals occurring in the dynamical equations, which have been verified to give no contribution \cite{Paper3,Paper4}. Moreover, the zero-range spin interaction, which represents a distinct feature of EC model, is absent; 
\item there exists a formal agreement between the radiative multipole moments of GR and EC theory. This will be proved  explicitly in this paper.
\end{enumerate}
Therefore, despite the profound differences between GR and EC frameworks, \emph{we have discovered, at 1PN order, some not a priori predictable formal similarities between the GR treatment of bodies endowed with a macroscopic angular momentum and the EC characterization of spinning objects modelled through the Weyssenhoff fluid}. However, it should be stressed that had we chosen an alternative fluid description, we might have attained distinct results. A scheme of our findings is given in Fig. \ref{fig:schema2}.  

Building upon the cited achievements, a wide variety of astrophysical applications involving both the dynamics and the ensuing radiative phenomena of either spinning  compact binaries or weakly self-gravitating spinning binaries framed either in GR or EC model is expected. The advantageous \emph{osmosis} between these two theories offers also the great opportunity to transfer a series of methodologies and results from one setting to the other. Exactly in this vein, in this work we aim to provide a set of analytical formulas regarding the orbital motion and the underlying coordinate time of binary systems whose spins are supposed to be aligned perpendicular to the orbital plane. These findings are extremely convenient for speeding up the calculations and avoiding numerical prescriptions, and hence can be timely employed to evaluate the two-body gravitational waveforms and fluxes. 

The paper is organized as follows. After having considered the dynamics and the radiative multipole moments of binary systems in Sec. \ref{sec:binary-system}, we derive the analytical formulas describing their relative orbit and coordinate time in Sec. \ref{sec:analytical-formulae}; in Sec. \ref{sec:applications}, we provide two applications, which involve the quantum spin and  the  macroscopic angular momentum of the bodies; finally, in Sec. \ref{sec:end} we draw the conclusions and outline future perspectives.

\subsection{Notations and conventions}

Greek indices take values  $0,1,2,3$, while lowercase Latin ones $1,2,3$. 
The spacetime coordinates are $x^\mu = (ct,\boldsymbol{x})$. Four-vectors are written as $a^\mu = (a^0,\boldsymbol{a})$, and $\boldsymbol{a} \cdot \boldsymbol{b}:= \delta_{lk}a^l b^k$, $\vert \boldsymbol{a} \vert\equiv a :=  \left(\boldsymbol{a} \cdot \boldsymbol{a}\right)^{1/2}$, and $\left(\boldsymbol{a} \times \boldsymbol{b}\right)^i := \varepsilon_{ilk} a^l b^k$, where $\varepsilon_{kli}$ is the total antisymmetric Levi-Civita symbol. The symmetric-trace-free (STF) projection of a tensor $A^{ij\dots k}$ is indicated with $A^{\langle ij\dots k \rangle }$. Superscripts $(l)$ denote $l$ successive time derivatives. $L=i_1 i_2 \dots i_l$ denotes a multi-index consisting of $l$ spatial indices. 

\section{Binary-system equations of motion and multipole moments}
\label{sec:binary-system}
In this section, we set out the dynamics and the radiative multipole moments of binary systems, which  can be obtained by applying the point-particle limit to a continuous fluid model \cite{Poisson-Will2014}. This procedure relies on the assumption that the fluid can be decomposed in a collection of $N=2$  mutually well separated components, each of them representing a body (of course, this approach can be easily generalized to any $N \geq 2$). This methodology entails the introduction of some center-of-mass variables which permit to substitute the fine-grained description of the system based on a number of fluid variables (such as the density and the pressure) with a coarse-grained picture. We have pursued this scheme in  Refs. \cite{Paper2,Paper3,Paper4}, where we have exploited 
the semiclassical Weyssenhoff model of a neutral spinning perfect fluid in EC theory and the PN formalism.

The section is then organized as follows.  We start with the 1PN equations of motion for binary systems and the resulting first integrals  (see Secs. \ref{Sec:equations of motion} and \ref{Sec:First-integrals}). Subsequently, after having displayed the general formulas of the waveform and the fluxes (cf. Sec. \ref{Sec:Grav-waveform-fluxes}), we derive the underlying radiative multipole moments and show their formal analogy with their GR counterpart in Sec. \ref{Sec:Rad-multipole-moments}. 

Henceforth, the bodies are labelled by  capital letters $A,B=1,2$.

\subsection{Equations of motion}
\label{Sec:equations of motion}
We consider a binary system composed of two spinning, weakly self-gravitating, slowly moving, and widely separated companions with masses $m_1\ge m_2$, total mass $M=m_1+m_2$, reduced mass $\mu=\tfrac{m_1m_2}{M}$, and symmetric mass ratio $\nu=\tfrac{\mu}{M}$. Given a harmonic coordinate system $x^\mu$, let $\boldsymbol{r}_A$ be the position vector,  $\boldsymbol{v}_A=\frac{{\rm d}\boldsymbol{r}_A}{{\rm d}t}$ the velocity, and $\boldsymbol{s}_A$ the spin of the objects. The latter is defined by
\begin{align} \label{spin-vector-def}
 \varepsilon_{jki}s_A^i(t) &:=\int_A {\rm d}^3 \boldsymbol{x} \, s_{jk}, 
\end{align}
where  $s_{\mu \nu}$ is the spin density tensor (cf. Eq. \eqref{spin-tensor-Weyssenhoff-fluid}). The above equation reflects the well-known  fact that in EC model  the quantum spin $\boldsymbol{s}_{A}$ is related to a geometrical feature of the spacetime, i.e., the torsion tensor. Furthermore, it makes   clear the difference between $\boldsymbol{s}_{A}$ and  the macroscopic angular momentum vector $\hat{\boldsymbol{s}}_{A}$  adopted in GR. In particular,  the former cannot be written in terms of kinematical quantities, unlike the latter, which is defined through an integral involving the density of the fluid and the cross product between the position and the velocity vectors of a fluid element relative to the center of mass (see Ref. \cite{Poisson-Will2014}, for more details). 

The motion of the binary system can be conveniently described by   choosing  an  orthogonal reference frame centered in the barycenter, which, without loss of generality, is supposed to be static.  In this frame, after  having  defined the relative  vectors  $\boldsymbol{R}:=\boldsymbol{r}_1-\boldsymbol{r}_2$ and $\boldsymbol{V}:= \tfrac{\dd}{\dd t}\boldsymbol{R}$, we find that the 1PN translational dynamics  is ruled by the relative acceleration
\begin{align}
\boldsymbol{A}:= \dfrac{\dd}{\dd t}\boldsymbol{V}= \boldsymbol{A}_{\rm GR} + \boldsymbol{A}_{\rm EC} + \OO \left(c^{-4}\right), 
\label{acceleration-EC-theory}
\end{align}
where \cite{Paper3}
\begin{subequations}
\label{A-GR-and-A-EC}
\begin{align}
\boldsymbol{A}_{\rm GR} &= -\frac{GM}{R^2} \boldsymbol{N} + \frac{GM}{c^2 R^2} \Biggl\{ \Bigl[ 2 (2 + \nu) \frac{GM}{R} + \frac{3\nu}{2}  \left(\boldsymbol{N} \cdot \boldsymbol{V}\right)^2 
\nn \\
&- (1+3 \nu) V^2 \Bigr] \boldsymbol{N} + 2(2-\nu) \left(\boldsymbol{N} \cdot \boldsymbol{V}\right) \boldsymbol{V} \Biggr\},
\label{GR-relative-acceleration}
\\
\boldsymbol{A}_{\rm EC} &=\frac{-4G}{c^2R^3} \Biggl[ \boldsymbol{V} \times \left(2 \boldsymbol{s} + \frac{3}{2}\boldsymbol{\sigma}\right) - 3 \boldsymbol{N} \left(\boldsymbol{N} \times \boldsymbol{V}\right) \cdot \left(\boldsymbol{s} + \boldsymbol{\sigma}\right)
\nn \\
&- 3 \boldsymbol{N} \times \left(\boldsymbol{s}  +\frac{\boldsymbol{\sigma}}{2}\right)\left(\boldsymbol{N} \cdot \boldsymbol{V}\right) \Biggr] -\frac{12G}{c^2 R^4 \mu} \Biggl\{ \boldsymbol{s}_1 \left(\boldsymbol{N}\cdot \boldsymbol{s}_2\right)
\nn \\
&+ \boldsymbol{s}_2 \left(\boldsymbol{N}\cdot \boldsymbol{s}_1\right) + \boldsymbol{N} \Bigl[ \boldsymbol{s}_1 \cdot \boldsymbol{s}_2 -5 \left(\boldsymbol{N}\cdot \boldsymbol{s}_1\right) \left(\boldsymbol{N}\cdot \boldsymbol{s}_2\right) \Bigr] \Biggl\},
\label{EC-relative-acceleration}
\end{align}
\end{subequations}
with   $\boldsymbol{N}= \boldsymbol{R}/R $  and 
\begin{align}
\boldsymbol{s}:= \boldsymbol{s}_1 + \boldsymbol{s}_2, \qquad \boldsymbol{\sigma}:= \dfrac{m_2}{m_1}\boldsymbol{s}_1 + \dfrac{m_1}{m_2}\boldsymbol{s}_2.
\end{align}
The 1PN motion is determined by Eqs. \eqref{acceleration-EC-theory} and \eqref{A-GR-and-A-EC} jointly with the conservation law $\dd \boldsymbol{s}_A /\dd t =\OO\left(c^{-2}\right) $  (see Ref. \cite{Paper3}, for further details).

\subsection{First integrals}
\label{Sec:First-integrals}

As shown in Ref. \cite{Paper4}, the 1PN dynamics of the binary system follows from an acceleration-dependent Lagrangian, which permits to determine the expressions of the related conserved energy and angular momentum. The total specific energy $E$ can be written as
\begin{equation} \label{conserved-energy}
E=E_{\rm GR}+E_{\rm EC} + \OO\left(c^{-4}\right),    
\end{equation}
where 
\begin{subequations} \label{eq:ENERGY}
\begin{align} 
E_{\rm GR}&=\left(\frac{V^2}{2}-\frac{G M}{R}\right) + \frac{1}{c^2}\Biggr{\{}\frac{G M}{2 R} \Biggl[\frac{G M}{R}+\nu  (\boldsymbol{N}\cdot\boldsymbol{V})^2   
\nonumber \\
&+(\nu +3) V^2\Biggr] +\frac{3}{8}(1-3 \nu ) V^4\Biggr{\}},
\\
E_{\rm EC}&=\frac{2G}{c^2 R^2}\Biggl\{ (\boldsymbol{N}\times\boldsymbol{V})\cdot \boldsymbol{\sigma} +\frac{2}{\mu R}\Biggr{[}3(\boldsymbol{N}\cdot\boldsymbol{s}_1)( \boldsymbol{N}\cdot\boldsymbol{s}_2)
\nonumber \\
&-\boldsymbol{s}_1\cdot\boldsymbol{s}_2\Biggr{]}\Biggr\},
\end{align}
\end{subequations}
while the total specific angular momentum $\boldsymbol{J}$  reads as
\begin{equation} \label{conserved-ang-mom}
\boldsymbol{J}=\boldsymbol{L}_{\rm GR}+\boldsymbol{L}_{\rm EC} +\frac{\bar{\boldsymbol{s}}}{\mu}+ \OO\left(c^{-4}\right),    
\end{equation}
with 
\begin{subequations} \label{eq:ANG_MOMENTUM}
\begin{align}
\boldsymbol{L}_{\rm GR}&= \boldsymbol{L}_{\rm N}\Biggl\{ 1 + \frac{1}{c^2} \left[\frac{G M}{R}(\nu +3)+\frac{(1-3 \nu )}{2}V^2\right]\Biggr\},\\
\boldsymbol{L}_{\rm EC}&=\frac{2 }{c^2 M}  \Biggr{\{}\frac{G M}{R} \, \boldsymbol{N}\times \left[\boldsymbol{N}\times (\boldsymbol{\sigma} +2 \boldsymbol{s} )\right]
\notag\\
&-\frac{1}{2} \boldsymbol{V}\times \left(\boldsymbol{V}\times \boldsymbol{\sigma} \right)\Biggr{\}}.
\end{align}
\end{subequations}
In the above equations, the Newtonian specific angular momentum is 
\begin{align}
    \boldsymbol{L}_{\rm N}= \boldsymbol{R} \times \boldsymbol{V},
\end{align}
and we have introduced the total spin vector  
\begin{align}
\bar{\boldsymbol{s}}:=\bar{\boldsymbol{s}}_1+\bar{\boldsymbol{s}}_2,    
\label{refined-spin-total}
\end{align}
where $\bar{\boldsymbol{s}}_{A}$  is the refined spin which keeps its magnitude constant during the motion (i.e., it satisfies $\bar{\boldsymbol{s}}_A \cdot \dd \bar{\boldsymbol{s}}_A / \dd t =0$).  It is  defined as \cite{Paper4}
\begin{align}
 \bar{\boldsymbol{s}}_A & := \boldsymbol{s}_A + \frac{1}{c^2} \left[\frac{G m_B}{R} \boldsymbol{s}_A +\frac{1}{2} \left(\boldsymbol{s}_A \cdot \boldsymbol{V}_A\right) \boldsymbol{V}_A\right] + \OC{-4},  
 \nonumber \\
 & \;  (A \neq B). 
 \label{refined-spin-s-bar}
\end{align}
As noted in Refs. \cite{Paper3,Paper4}, both the translational and the rotational 1PN GR dynamics pertaining to \emph{weakly self-gravitating} binary systems  having  a macroscopic angular momentum vector $\hat{\boldsymbol{s}}_A$ are  formally recovered if the substitution
\begin{align} \label{spin-and-hat-spin}
\boldsymbol{s}_A \rightarrow \frac{1}{2}   \hat{\boldsymbol{s}}_A, 
\end{align}
is applied to the 1PN motion in EC theory. As it will be clear from our forthcoming analysis, in the case of maximally rotating \emph{compact objects}, the above relation should be slightly modified according to 
\begin{align} \label{spin-and-hat-spin-2}
 \boldsymbol{s}_A \rightarrow \frac{1}{2c}   \hat{\boldsymbol{s}}_A,  
\end{align}
where, following the usual GR conventions \cite{Blanchet2006a}, now $ \hat{\boldsymbol{s}}_A$  has the dimensions of an angular momentum multiplied by $c$. Note that, in order to ease the notations,  we will always use the symbol  $\hat{\boldsymbol{s}}_A$, although the physical dimensions of the macroscopic angular momentum are different in Eqs. \eqref{spin-and-hat-spin} and \eqref{spin-and-hat-spin-2}. Indeed, this should not create any confusion, as it will be clear from the context to which relation we will refer.

\subsection{Gravitational waveform and fluxes}
\label{Sec:Grav-waveform-fluxes}

Let us indicate with $I^{\rm rad}_L$ and  $J^{\rm rad}_L$ the STF mass-type and  current-type radiative multipole moments of order $l$, respectively.  Bearing in mind that, at 1PN level, there is no  difference between the harmonic  and the radiative coordinates \cite{Blanchet-Damour1989,Blanchet2014},  the 1PN-accurate \emph{asymptotic waveform} $\mathscr{H}^{\rm TT}_{ij}$   reads as \cite{Thorne1980,Paper1,Paper2}
\begin{align} \label{gravitational_wave_amplitude}
    \mathscr{H}_{ij}^{\rm TT}(x^\mu) & = \dfrac{2G}{c^4 \vert\boldsymbol{x}\vert} \mathscr{P}_{ijkl}(\boldsymbol{n})  \Biggr\{ \overset{(2)}{I}{}_{kl}^{{\rm rad}}(u) 
    \nonumber \\
    & + \dfrac{1}{c} \left[  \dfrac{1}{3} n_a  \overset{(3)}{I}{}_{kla}^{{\rm rad}}(u) +\dfrac{4}{3} n_b \epsilon_{ab(k} \overset{(2)}{J}{}_{l)a}^{{\rm rad}}(u)  \right]
    \nonumber \\
    & +\dfrac{1}{c^2} \left[\dfrac{1}{12}n_{a}n_{b}   \overset{(4)}{I}{}_{klab}^{{\rm rad}}(u) \right.
    \nonumber \\
     &\left. + \dfrac{1}{2}n_{b}n_{c} \epsilon_{ab(k}   \overset{(3)}{J}{}_{l)ac}^{{\rm rad}}(u)  \right]
      + {\rm O}(c^{-3}) \Biggr\}, 
\end{align}
where $u= t-\vert\boldsymbol{x}\vert/c$,  $\boldsymbol{n}=\boldsymbol{x}/\vert\boldsymbol{x}\vert$, and $\mathscr{P}_{ijkl}(\boldsymbol{n})$ is transverse-traceless (TT) projection operator onto the plane orthogonal to $\boldsymbol{n}$.  Moreover, the  \textit{total radiated power} $\mathcal{F}$ (also dubbed \textit{energy flux} or \textit{gravitational luminosity}) and   the \emph{angular momentum flux} $\mathcal{G}_i$ read as \cite{Thorne1980} 
\begin{subequations}
\label{fluxes}
\begin{align} 
\mathcal{F}(t) & = \dfrac{G}{c^5} \Biggl\{\dfrac{1}{5}\overset{(3)}{I}{}_{ij}^{{\rm rad}}\overset{(3)}{I}{}^{{\rm rad}}_{ij} + \dfrac{1}{c^2} \Biggl[ \dfrac{1}{189} \overset{(4)}{I}{}^{{\rm rad}}_{ijk} \overset{(4)}{I}{}^{{\rm rad}}_{ijk}
 \nonumber \\
&+ \dfrac{16}{45} \overset{(3)}{J}{}^{\rm rad}_{ij}\overset{(3)}{J}{}_{ij}^{\rm rad} \Biggr]
+ {\rm O}\left(c^{-4}\right) \Biggr \}, 
\label{power-radiated-1PN} \\
\mathcal{G}_i(t)   &= \dfrac{G}{c^5} \varepsilon_{ijk} \Biggl\{\dfrac{2}{5} \overset{(2)}{I}{}^{{\rm rad}}_{jl}\overset{(3)}{I}{}^{{\rm rad}}_{kl} 
   + \dfrac{1}{c^2} \Biggl[ \dfrac{1}{63} \overset{(3)}{I}{}^{{\rm rad}}_{jlp} \overset{(4)}{I}{}^{{\rm rad}}_{klp}
 \nonumber \\
&+ \dfrac{32}{45} \overset{(2)}{J}{}^{{\rm rad}}_{jl}\overset{(3)}{J}{}^{{\rm rad}}_{kl} \Biggr]
+ {\rm O}\left(c^{-4}\right) \Biggr \}. \label{ang-mom-flux-1PN}
\end{align}
\end{subequations}

Note that the linear momentum flux involves a Newtonian formula (see Ref. \cite{Thorne1980}, for more details)  and hence  will not be considered in this paper. 

\subsection{Radiative multipole moments}
\label{Sec:Rad-multipole-moments}
The general form, with the required PN accuracy, of the radiative multipole moments occurring in the formulas of Sec. \ref{Sec:Grav-waveform-fluxes}  has been first obtained in Ref. \cite{Paper1}, where we have solved the GW generation problem in EC theory via the Blanchet-Damour formalism\footnote{The  radiative moments $I^{\rm rad}_L, J^{\rm rad}_L$ are related to the  STF  radiative multipole moments $U_L, V_L$ employed in Ref. \cite{Paper1} by the relations $U_L := \overset{(l)}{I}{}_{L}^{{\rm rad}}, V_L :=\overset{(l)}{J}{}_{L}^{{\rm rad}}$.}. Then, the expressions valid in the case of spinning binary systems have been  derived in Ref. \cite{Paper2}. Starting from these  results and adopting a mass-centered coordinate system, we  find that, after some manipulations,  the mass-type radiative moments can be written as 
\begin{align}
I^{\rm rad}_{ij}& =\mu R_{\langle ij\rangle }\left[1+\frac{29}{42c^2}(1-3\nu)V^2-\frac{(5-8\nu)}{7c^2}\frac{GM}{R}\right]\notag\\
&+\frac{\mu(1-3\nu)}{21c^2}\left[11R^2 V_{\langle ij\rangle } -12 (\boldsymbol{R} \cdot \boldsymbol{V}) R_{\langle i} V_{j \rangle }\right]\notag\\
&+\frac{8\nu}{3c^2}\left[2(\boldsymbol{V}\times\boldsymbol{\sigma})^{\langle i}R^{j\rangle }  -\left( \boldsymbol{R}\times\boldsymbol{\sigma}\right)^{\langle i}V^{j \rangle }\right]\notag\\
&+{\rm O}\left(c^{-3}\right),
\label{I-ij-rad-pp-limit-expression-2}\\
I^{\rm rad}_{ijk}& =-\mu\sqrt{1-4\nu}R_{\langle ijk\rangle }+{\rm O}\left(c^{-2}\right),
\\
I^{\rm rad}_{ijkl}& =\mu(1-3\nu)R_{\langle ijkl\rangle} +{\rm O}\left(c^{-2}\right),
\end{align}
while   the current-type radiative moments are
\begin{align}
J^{\rm rad}_{ij}& =-\mu\sqrt{1-4\nu}\epsilon_{kl\langle i}R_{j\rangle k}V_l\notag\\
&+3\mu\left(\frac{s_1^{\langle i}R^{j\rangle }}{m_1}-\frac{s_2^{\langle i}R^{j\rangle }}{m_2}\right)
+{\rm O}\left(c^{-2}\right),
\label{J-ij-rad-pp-limit-expression-2}
\\
J^{\rm rad}_{ijk}&=\mu(1-3\nu)R_{\langle ij}\epsilon_{k\rangle lp}R_l V_p +4\nu  R^{\langle i} R^j \sigma^{k \rangle}+{\rm O}\left(c^{-2}\right).
\label{J-ijk-rad-pp-limit-expression-2}
\end{align}
Notice that in Eq. \eqref{I-ij-rad-pp-limit-expression-2} we have  exploited the equations of motion \eqref{acceleration-EC-theory} jointly with the fact that  the spin vector is conserved modulo ${\rm O}\left(c^{-2}\right)$ corrections. 

At this stage, some comments are in order. First of all,  Eqs. \eqref{I-ij-rad-pp-limit-expression-2}, \eqref{J-ij-rad-pp-limit-expression-2}, and \eqref{J-ijk-rad-pp-limit-expression-2} agree formally with their corresponding GR moments \cite{Kidder1993,Kidder1995,Blanchet2014} if the quantum spin  $\boldsymbol{s}_{A}$ is replaced by the macroscopic angular momentum vector $\hat{\boldsymbol{s}}_{A}$  following the   scheme already introduced  in Eqs. \eqref{spin-and-hat-spin} and \eqref{spin-and-hat-spin-2}. This is a crucial consistency check, since, as pointed out before, the same conclusion is valid in the context of  the 1PN dynamics of spinning binaries. Moreover, this is a remarkable result if we recall that Eqs. \eqref{I-ij-rad-pp-limit-expression-2}--\eqref{J-ijk-rad-pp-limit-expression-2} have been obtained  by applying the point-particle procedure to the radiative moments pertaining to a Weyssenhoff fluid in EC theory, i.e., a generalization of the usual perfect fluid adopted in GR. Furthermore, we see that in our framework Eqs.  \eqref{I-ij-rad-pp-limit-expression-2}--\eqref{J-ijk-rad-pp-limit-expression-2} are immediately consistent with the Frenkel spin supplementary condition. On the other hand, in the context of GR the radiative moments follow the center-of-mass definition stemming from the Frenkel constraint only if a suitable transformation is invoked (see Eq. (13) in Ref. \cite{Kidder1993} and Appendix A in Ref. \cite{Kidder1995}, for further details). 

\section{Analytical formulas}
\label{sec:analytical-formulae}
In this section, we investigate the dynamical and radiative features of binary systems analytically. We suppose that the spins of the companions and the orbital angular momentum $\boldsymbol{L}:=\boldsymbol{L}_{\rm GR}+\boldsymbol{L}_{\rm EC}$ are aligned, namely (cf. Eqs. \eqref{conserved-ang-mom}--\eqref{refined-spin-s-bar}) 
\begin{align}\label{aligned-spin}
\boldsymbol{\bar s}\cdot\boldsymbol{R}=0, \quad \boldsymbol{\bar s}\cdot\boldsymbol{V}=0, \quad   \boldsymbol{\bar s}_1\times\boldsymbol{\bar s}_2=\boldsymbol{0}.     
\end{align}
In this setting, there is no spin precession, as $\dd \bar{\boldsymbol{s}}_A/\dd t=\OC{-4}$ (cf. Eqs. (27) and (28) in Ref. \cite{Paper4}) and the  dynamics takes place in a fixed plane (see Eqs. \eqref{acceleration-EC-theory} and \eqref{A-GR-and-A-EC}). Thus, we can set the barycentric frame $(x,y,z)$ in such a way that the motion occurs in the plane $(x,y)$, where we introduce polar coordinates $(R,\theta)$, $\theta$ being the angle between $\boldsymbol{R}$ and the $x$-axis measured counterclockwise. This also means that we can write $\bar{\boldsymbol{s}}_1=(0,0,\bar{s}_{1z})$ and  $\bar{\boldsymbol{s}}_2=(0,0,\bar{s}_{2z})$. 

In our hypotheses, the orbital angular momentum  $\boldsymbol{L}$ and the total spin vector $\bar{\boldsymbol{s}}$ are \emph{separately} conserved modulo $\OC{-4}$ corrections (see Eq. \eqref{conserved-ang-mom}). In other words, we have one additional first integral with respect to the most general case. We can take advantage of this situation, as the number of conserved quantities and degrees of freedom of the system coincide. Therefore, the conserved $\bar{\boldsymbol{s}}$ resolves the rotational motion as it gives the only nonvanishing component of the spins, whereas the constants $E$ and $L$ determine the translational motion, as they can be used to work out the functions $R(t)$ and $\theta(t)$. In this configuration, we are able to parametrize  the relative orbit \emph{\'a la Damour-Deruelle} (see Sec. \ref{sec:relative_orbit}). Here, the coordinate time plays a key role and its  analytical expression is worked out in Sec. \ref{sec.coordinate_time}. Such  formulas are essential to evaluate the energy and angular momentum fluxes. These are given, for generic spin orientation, in Sec. \ref{Sec:Analytic-energy-and-ang-mom-fluxes}, where we also derive some new contributions, which, to the best of our knowledge, have not been  transcribed in the literature.

\subsection{Polar equation of the relative orbit}
\label{sec:relative_orbit}

Starting from the expressions of $E$ and $L$, and employing polar coordinates, we find that the 1PN relative dynamics of the companions is described by
\begin{subequations} \label{eqs_EC_split}
\begin{align}
\left(\frac{\dd R}{\dd t}\right)^2&=\mathcal{A}+\frac{2\mathcal{B}}{R}+\frac{\mathcal{C}}{R^2}+\frac{\mathcal{D}}{R^3} + \OC{-4},\\
\frac{\dd \theta}{\dd t}&=\frac{\mathcal{H}}{R^2}+\frac{\mathcal{I}}{R^3} + \OC{-4},\label{eq:angle_eq}
\end{align}
\end{subequations}
with 
\begin{subequations} 
\label{coefficients-DD-EC}
\begin{align}
\mathcal{A}&=A,\\    
\mathcal{B}&=B,\\
\mathcal{C}&=C+\frac{1}{c^2}\frac{4LE}{M}\sigma_z,\\
\mathcal{D}&=D+\frac{4G}{c^2}\left[2\frac{s_{1z}s_{2z}}{\mu}-L(2\sigma_z+s_z)\right],\\
\mathcal{H}&=H-\frac{1}{c^2}\frac{2E}{M}\sigma_z,\\
\mathcal{I}&=I+\frac{4G}{c^2}s_z,
\end{align} 
\end{subequations}
where $s_z=s_{1z}+s_{2z}$, $\sigma_z=\frac{m_2}{m_1}s_{1z}+\frac{m_1}{m_2}s_{2z}$,  $\mathfrak{s}_z :=s_{1z}s_{2z}/\mu$ (recall that $\bar{\boldsymbol{s}}_A = \boldsymbol{s}_A + \OC{-2}$, cf. Eq. \eqref{refined-spin-s-bar}). The parameters $A,B,C,D,H,I$ can be found in the  Damour and Deruelle paper (see Eq. (2.17) in Ref. \cite{Damour1985}) with the caveat that the energy $E$ and the orbital angular momentum $L$ must be read off from our Eqs. \eqref{conserved-energy}--\eqref{eq:ANG_MOMENTUM}.

Equation \eqref{eqs_EC_split} assumes the same functional form as in GR  (see Eqs. (2.15) and (2.16) in Ref. \cite{Damour1985}). We can then employ the strategy pursued by Damour and Deruelle, which uses conchoidal transformations in order to map  Eq. \eqref{eqs_EC_split} to an auxiliary Newtonian-like form. It should be clear that, in our case, these transformations mirror formally those of GR, the only difference being the occurrence  of (some of) the coefficients displayed in Eq.  \eqref{coefficients-DD-EC}. Therefore, we can parametrize the radial and the angular 1PN motion in EC model as
\begin{subequations} \label{orbit_parametrization_GR}
\begin{align}
&n(t-t_0)=u-e_t\sin u + \OC{-4},\\
&R(t)=a_R(1-e_R \cos u) + \OC{-4},\\
&\theta(t)=\theta_0+2K\arctan\left[\left(\frac{1+e_\theta}{1-e_\theta}\right)\tan \frac{u}{2}\right] + \OC{-4},
\end{align}    
\end{subequations}
where the 1PN quantities $\{n,a_R,e_R,e_t,e_\theta,K\}$ are 
\begin{subequations} \label{PN_parameters_GR}
\begin{align}
n&=\frac{(-\mathcal{A})^{\frac{3}{2}}}{\mathcal{B}},\\
a_R&=-\frac{\mathcal{B}}{\mathcal{A}}-\frac{\mathcal{D}}{2L^2},\\
e_t&=\left[1-\frac{\mathcal{A}}{\mathcal{B}^2}\left(\mathcal{C}+\frac{\mathcal{B}\mathcal{D}}{L^2}\right)\right]^{1/2},\\
e_R&=\left(1-\frac{\mathcal{A}\mathcal{D}}{2\mathcal{B}L^2}\right)e_t,\\
e_\theta&=\left(1-\frac{\mathcal{A}\mathcal{D}}{\mathcal{B}L^2}-\frac{\mathcal{A}\mathcal{I}}{\mathcal{B}\mathcal{H}}\right)e_t,\\
K&=1+\frac{1}{c^2}\frac{3G^2M^2}{L^2}.
\label{parameter-K}
\end{align}    
\end{subequations}

Starting from the above relations, it is easy to show that the polar equation of the 1PN relative orbit is
\begin{align} \label{eq:radius}
R(\theta)&=\frac{e_R}{e_\theta}a_R\frac{1-e_\theta^2}{1+e_\theta \cos\left(\frac{\theta-\theta_0}{K}\right)}+a_R\left(1-\frac{e_R}{e_\theta}\right) 
\nonumber \\
&+ \OC{-4}.   
\end{align}

\subsection{Coordinate time} 
\label{sec.coordinate_time}
Having obtained the orbital radius \eqref{eq:radius}, we can now derive the analytical formula of $t(\theta)$, which expresses  the coordinate time  as a function of the polar angle. This can be done by generalizing  the strategy developed in Ref. \cite{LetteraDBA}, which relies on first determining a main discontinuous function $f(\theta)$, which is then made smooth via the introduction of the accumulation function $F_{n}(\theta)$. 

In order to build up the differential equation for $t(\theta)$, we split the energy $E$, the orbital angular momentum $L$, and the radius $R$ as follows: $E=E_0+\frac{1}{c^2}E_1 +\OC{-4}$,  $L=L_0+\frac{1}{c^2}L_1 +\OC{-4}$, and $R=R_0+\frac{1}{c^2}R_1+\OC{-4}$. The expressions of $E_0,E_1,L_0$, and $L_1$ can be promptly read off from Eqs. \eqref{eq:ENERGY} and \eqref{eq:ANG_MOMENTUM}, whereas for $R_0$ and $R_1$ we make use of Eq. \eqref{eq:radius}. Since  EC theory is the same as  GR at  0PN level,  $R_0$ reads as (cf. Eq. (5a) in Ref. \cite{LetteraDBA})
\begin{align}
R_0&=\frac{1}{B_1+B_2 \cos (\tilde{K}\tilde{\theta})},
\label{eq:R0} 
\end{align}
where $\tilde{\theta}=\theta -\theta_0$, $B_1> B_2\ge 0$, and 
\begin{subequations}
\begin{align}
B_1&= \frac{1}{h_0^2 GM},\\
B_2&= e_0 B_1,\\
\tilde{K}&=1-\frac{1}{c^2}\frac{3}{h_0^2}+\frac{2 G}{c^2 h_0L_0^2}\left[4s_z+3\sigma_z-6\frac{\mathfrak{s}_z}{L_0}\right],\\
h_0&= \frac{L_0}{GM},
\\
e_0&=\sqrt{1+2E_0 h_0^2}, \qquad (0\le e_0 < 1),
\end{align}
\end{subequations} 
where $e_0$ is the Newtonian eccentricity and  $\tilde{K}$  the 1PN expansion of $1/K$ (cf. Eq. \eqref{parameter-K}). This last factor is responsible for the orbit precession, which is influenced also by the presence  of the spin. The term  $R_1$ can be written as the sum of the GR and EC contributions, i.e., $R_1=R_1^{\rm GR}+R_1^{\rm EC}$. We find (cf. Eq. (5b) in Ref. \cite{LetteraDBA})
\begin{subequations} \label{eq:R1}
\begin{align}
R_1^{\rm GR}&=
A_0+A_1 R_0+A_2 R_0^2 \cos(\tilde{K}\tilde{\theta}), \label{eq:R1_GR}\\
R_1^{\rm EC}&=R_0^2\left[W_1+W_2\cos(\tilde{K}\tilde{\theta})+W_3\cos(\tilde{K}\tilde{\theta})\right],
\end{align}
\end{subequations}
where 
\begin{subequations}
\begin{align}
A_0&=\frac{G \mu }{2},\\
A_1&=2 E_0 \left(\frac{\nu }{2}-2\right)+W_0,\\
A_2&=-\frac{E_0^2}{e_0 G M} \left(\frac{E_1}{E_0^2}+\frac{2 W_0}{E_0}+\frac{\nu -15}{2}\right),\\
W_1&=\frac{4 \left(e_0^2+3\right) s_z+\left(e_0^2+8\right)\sigma_z}{G^2M^3 h_0^5}-\frac{6 \left(e_0^2+2\right) \mathfrak{s}_z}{G^3 h_0^6 M^4},\\
W_2&=\frac{4 \left(3 e_0^2+1\right) (GMh_0 s_z-\mathfrak{s}_z)}{G^3M^4h_0^6e_0}\notag\\
&+\frac{\left(3+6 e_0^2-e_0^4\right) GMh_0  \sigma_z}{e_0 G^3M^4h_0^6},\\
W_3&=\frac{e_0^2 (2 \mathfrak{s}_z-G h_0 M \sigma_z)}{G^3M^4h_0^6},
\end{align}
\end{subequations} 
and $W_0=L_1/L_0-3/h_0^2$. 

If we write the 1PN differential equation for $t(\theta)$  in the form (see Eq. \eqref{eq:angle_eq})  
\begin{align} \label{eq:Time_original}
{\rm d}t&=\frac{{\rm d}\theta}{\frac{\mathcal{H}}{R^2}+\frac{\mathcal{I}}{R^3}}+\OC{-4}=\dd t_{\rm GR}+\dd t_{\rm EC}+\OC{-4},
\end{align}
then the substitution of the splittings shown above for $E,L$, and $R$ gives (cf. Eq. (8) in Ref. \cite{LetteraDBA})
\begin{subequations}
\begin{align}
\dd t_{\rm GR}&=\frac{R_0^2}{L_0}\biggr{\{}1+\frac{1}{c^2 }\biggr{[} E_0(1-3 \nu ) +\frac{2R_1}{R_0}\notag\\
&\qquad\qquad- \frac{4GM}{R_0}(\nu -2)-\frac{L_1}{L_0}\biggr{]}\biggr{\}}
\dd\theta,\\
\dd t_{\rm EC}&=\frac{2R_0 (E_0 R_0\sigma_z-2 G M s_z)}{c^2 L_0^2 M}\dd\theta.
\end{align}    
\end{subequations}
At this stage,  we can use the explicit formulas of $R_0$  and $R_1$ (cf. Eqs. \eqref{eq:R0} and  \eqref{eq:R1}) to obtain the complete expression of the differential equation for $t(\theta)$, which can be easily integrated yielding the result
\begin{align}
f(\theta)&=\frac{1}{c^2 L_0^2 \tilde{K}}\Biggr{\{}(C_1 R_0+C_2)R_0 \sin (\tilde{K} \tilde{\theta})\notag\\
&+C_0 \arctan \left[\sqrt{\frac{B_1-B_2}{B_1+B_2}} \tan \left(\frac{\tilde{K}\tilde{\theta}}{2}\right)\right]
\Biggr{\}},
\end{align}
where 
\begin{subequations}
\begin{align}
C_0&=\frac{2 L_0}{\left(B_1^2-B_2^2\right)^{5/2}} \Biggr{\{}2 \left(B_1^2-B_2^2\right)^2 [A_0-G (\nu -2) M]\notag\\
&+B_1 \left(B_1^2-B_2^2\right) \left[\left(2 A_1+c^2-3 E_0 \nu +E_0\right)-\frac{L_1}{L_0}\right]\notag\\
&+3 B_2 [B_2 W_3-B_1 (A_2+W_2)]+W_1 \left(2 B_1^2+B_2^2\right)\Biggr{\}},\\
C_1&=\frac{L_0}{B_2 \left(B_1^2-B_2^2\right)} \Biggr{[}B_1 B_2 (A_2+W_2)-2 B_1^2 W_3\notag\\
&+B_2^2 (W_3-W_1)\Biggr{]},\\
C_2&=\frac{L_0}{B_2 \left(B_1^2-B_2^2\right)^2} \Biggr{\{}B_2^2 \left(B_2^2-B_1^2\right) \Biggr{[}\left(2 A_1+c^2\right.\notag\\
&\left.-3 E_0 \nu+E_0\right)-\frac{L_1}{L_0}\Biggr{]}+A_2 B_2 \left(B_1^2+2 B_2^2\right)+\left(2 B_1^3 W_3\right.\notag\\
&\left.+B_1^2 B_2 W_2-B_1 B_2^2 (3 W_1+5 W_3)+2 B_2^3 W_2\right)\Biggr{\}}.
\end{align}    
\end{subequations}
The function $f(\theta)$ is  discontinuous on intervals lying outside $[0,2\pi]$. Therefore, in order to regularly connect its different curve branches, we make use of  the accumulation function $F_n(\theta)$ \cite{LetteraDBA}, which reads as
\begin{align} \label{eq:acc_func}
F_n(\theta)=\begin{cases}
0, & {\rm if}\ \tilde{\theta}\in[0,P_\theta],\\
2n f(P_\theta), & {\rm if}\ \tilde{\theta}\in[P_\theta(2n+1),P_\theta(2n+2)],
\end{cases}
\end{align}
where $P_\theta=\pi/\tilde{K}$ is  the \emph{characteristic period} and $n\in \mathbb{N}$. For a generic $\theta$, the related value of $n$ can be calculated considering $q= [(\tilde{\theta}-P_\theta)/P_\theta]$, where $[\cdot]$ stands for the integer part of a number. Thus, if $q$ is an even number, then $n=(q+2)/2$; on the other hand, if $q$ is an odd number, then $n=(q+1)/2$. Therefore, we can conclude that the correct \emph{analytical} form of $t(\theta)$ is
\begin{align} \label{eq:anal_sol}
t(\theta)=f(\theta)+F_n(\theta) +{\rm O}\left(c^{-4}\right).   
\end{align}

\begin{figure}
\centering
\includegraphics[scale=0.29]{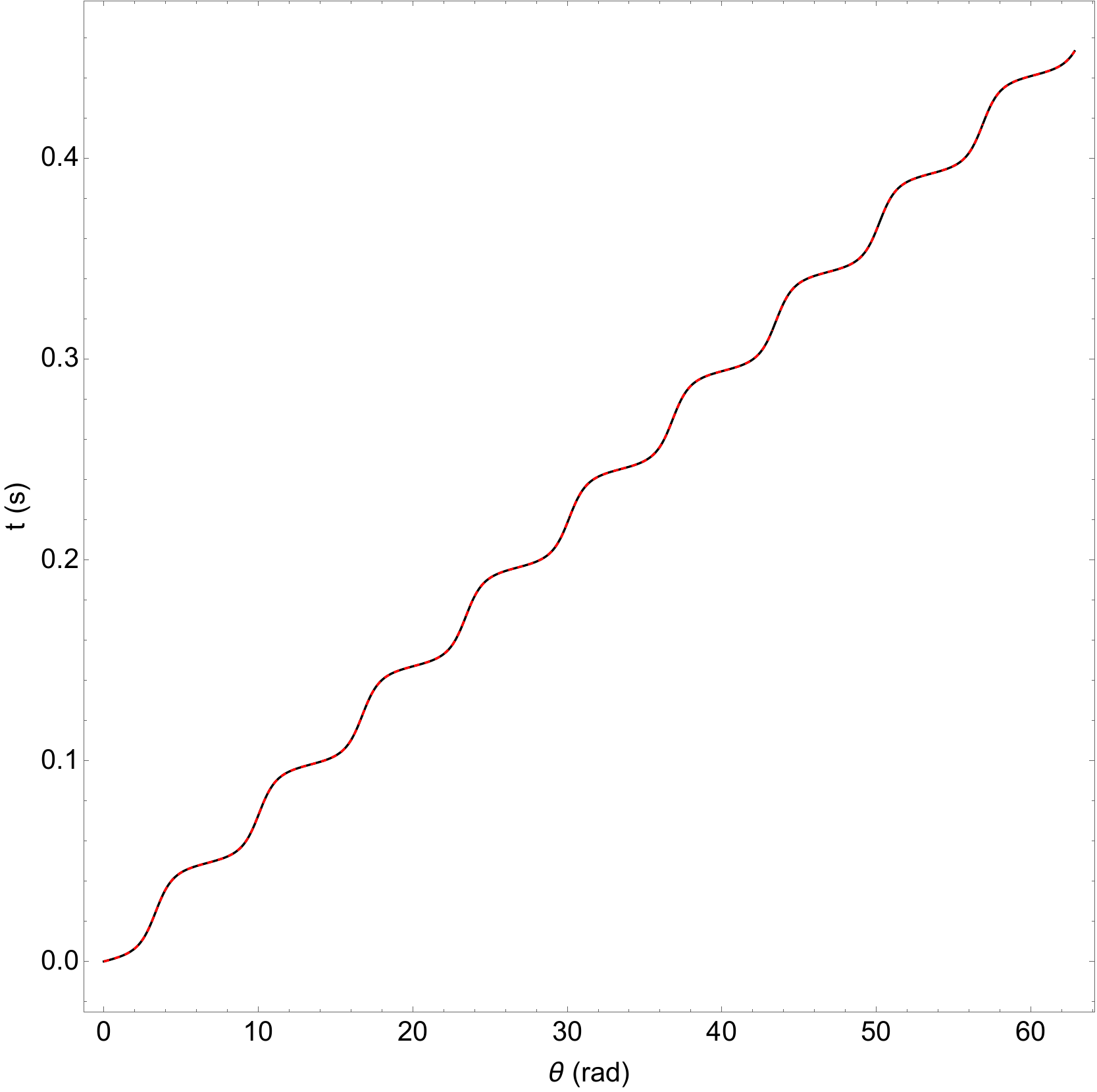}
\caption{Function $t(\theta)$ with $\theta\in[0,20\pi]$ for  a binary NS system having 
 the following parameters: $m_1=1.60\ M_\odot$, $m_2=1.17\ M_\odot$,  $\theta_0= 0$,  $E_0=-6.80\times 10^{14}\ {\rm m}^2  \ {\rm s}^{-2}$, $E_1=1.07\times 10^{30}$, $L_0=8.62\times 10^{12}\ {\rm  m^2\ s^{-1}}$, $L_1=2.57\times 10^{28}\ {\rm  s}$, $s_{1z}=1.21\times10^{57}\hbar$, $s_{2z}=4.73\times10^{56}\hbar$. The black continuous line represents the numerical solution, whereas the red dashed line the analytical expression \eqref{eq:anal_sol}.}
\label{fig:FigA3}
\end{figure}

In Fig. \ref{fig:FigA3}, we show the agreement between the numerical solution of Eq. \eqref{eq:Time_original} and the analytical expression \eqref{eq:anal_sol}. Here, we have considered two neutron stars (NSs), whose quantum spins are modeled as follows 
\begin{align}
\label{szi-components-NS}
s_{Az}=
\mathcal{N}\hbar\frac{4\pi}{3}\left(\frac{6Gm_A}{c^2}\right)^3,
\end{align}
where $\mathcal{N}=10^{44}\ \mbox{m}^{-3}$ is estimated as the inverse of the nucleon volume \cite{Paper1,Paper2}.  

The analytical expression of $t(\theta)$ is extremely useful for  speeding up the computations in several astrophysical applications, like (see Ref. \cite{LetteraDBA} and references therein): pulsar timing software such as \texttt{TEMPO2};  coherent pulsar search algorithms; GW astronomy, where it can be used to match the observational data with theoretical templates. 

\subsection{Energy and angular momentum fluxes}
\label{Sec:Analytic-energy-and-ang-mom-fluxes}

The analytic formulas presented in the previous sections play a fundamental role in the evaluation of the energy and angular momentum fluxes. Let us start with  the  luminosity \eqref{power-radiated-1PN}, which, for generic spin direction, can be written as
\begin{align} \label{flux_luminosity}
    \mathcal{F}&= \frac{8}{15}\frac{G^3 \mu^2M^2}{ c^5 R^4}\biggl[ \mathcal{F}_{\rm N}+\frac{1}{c^2}\bigl(\mathcal{F}_{\rm 1PN} \notag\\
    &+  \mathcal{F}_{\rm SO}+\mathcal{F}_{\rm SS}+\mathcal{F}_{\rm SS'}\bigr) + \OC{-3}\biggr].
\end{align}
Here,  the GR contributions are \cite{Blanchet-Schafer1989} 
\begin{subequations}
\begin{align}
\mathcal{F}_{\rm N} &=12V^2-11(\boldsymbol{N}\cdot\boldsymbol{V})^2,\\
\mathcal{F}_{\rm 1PN} &=\frac{G M}{7R}\left[\left(734-30\nu\right)  (\boldsymbol{N}\cdot\boldsymbol{V})^2+\left(4-16\nu\right) \frac{G M}{R}\right]\notag\\
&+\frac{V^2}{14}\biggr{[}\frac{80G M}{R}\left(\nu-17\right) + \left(\frac{785}{2}-426 \nu\right) V^2\notag\\
&+\left(1392 \nu-1487\right)(\boldsymbol{N}\cdot\boldsymbol{V})^2\biggr{]}+\frac{1}{7}\left(\frac{2061}{4}-465\nu\right)\notag\\
& \times(\boldsymbol{N}\cdot\boldsymbol{V})^4;
\end{align}
\end{subequations}
whereas for the EC corrections we find
\begin{subequations}
\begin{align}
\mathcal{F}_{\rm SO} &=\frac{2}{MR^2} \boldsymbol{L}_{\rm N} \cdot \left\{\boldsymbol{s}\left[78 (\boldsymbol{N}\cdot\boldsymbol{V})^2-\frac{8GM}{R}-80 V^2\right]\right.\notag\\
&\left.+\frac{\left(\boldsymbol{\chi}_2-\boldsymbol{\chi}_1\right)}{(m_1-m_2)^{-1}}\biggr{[}51 (\boldsymbol{N}\cdot\boldsymbol{V})^2+\frac{4 G M}{R}-43 V^2\biggr{]}\right\},\\
\mathcal{F}_{\rm SS} &=\frac{2}{R^2}\biggr{\{}3 (\boldsymbol{\chi}_1\cdot\boldsymbol{\chi}_2) \left[47 V^2-55 (\boldsymbol{N}\cdot\boldsymbol{V})^2\right]\notag\\
&-3 ( \boldsymbol{N}\cdot\boldsymbol{\chi}_1) ( \boldsymbol{N}\cdot\boldsymbol{\chi}_2) \left[168 V^2-269 (\boldsymbol{N}\cdot\boldsymbol{V})^2\right]\notag\\
&-17 (\boldsymbol{N}\cdot\boldsymbol{V}) \left[(\boldsymbol{N}\cdot\boldsymbol{\chi}_2) ( \boldsymbol{V}\cdot\boldsymbol{\chi}_1)\right.\notag\\
&\left. +(\boldsymbol{N}\cdot\boldsymbol{\chi}_1) (\boldsymbol{N}\cdot\boldsymbol{\chi}_2)\right]+71 (\boldsymbol{V}\cdot\boldsymbol{\chi}_1) ( \boldsymbol{V}\cdot\boldsymbol{\chi}_2)\biggr{\}},
\\
\mathcal{F}_{\rm SS'}&=\frac{1}{R^2}\sum_A\biggr{\{}3(\boldsymbol{\chi}_A\cdot\boldsymbol{\chi}_A)\biggr{[}3(\boldsymbol{N}\cdot\boldsymbol{V})^2+V^2\biggr{]}\notag\\
&+\biggr{[}3(\boldsymbol{N}\cdot\boldsymbol{V})(\boldsymbol{\chi}_A\cdot \boldsymbol{N})-(\boldsymbol{\chi}_A\cdot \boldsymbol{V})\biggr{]}^2\biggr{\}},
\label{Flux-new}
\end{align}
\end{subequations}
where we have defined
\begin{align}
\boldsymbol{\chi}_A&:=\frac{\boldsymbol{s}_A}{m_A}.
\end{align}

Some fundamental remarks on the above relations should now be given. First of all, we note that Eq. \eqref{Flux-new} contains a  spin-spin interaction going like $s_A^2$, which stems from $\overset{(3)}{J}{}^{\rm rad}_{ij}\overset{(3)}{J}{}_{ij}^{\rm rad}$ (while similar corrections due to $\overset{(3)}{I}{}_{ij}^{{\rm rad}}\overset{(3)}{I}{}^{{\rm rad}}_{ij}$ will arise at higher PN orders).  Such terms  occur also in GR, but they are not reported in Refs. \cite{Kidder1993,Kidder1995}\footnote{L. E. Kidder has informed us in a private communication that he has calculated these terms in GR. However, the final result was not published, since he deemed that the contributions proportional to $s_A^2$ should be combined with the corrections due to the quadrupole-moment tensor induced by the the oblateness of the bodies (see Ref. \cite{Poisson1997}, for more details).}. Therefore, to the best of our knowledge, formula \eqref{Flux-new} is displayed for the first time in this paper. Moreover, we stress again that Eq. \eqref{flux_luminosity} is formally analogous to the GR flux \cite{Kidder1993,Kidder1995,Blanchet2006b} if we consider either the substitution \eqref{spin-and-hat-spin} (in the case of weakly self-gravitating bodies) or  Eq. \eqref{spin-and-hat-spin-2} (for compact objects). Note that in EC model we consider that the quantities $\mathcal{F}_{\rm SO}, \mathcal{F}_{\rm SS}$, and $ \mathcal{F}_{\rm SS'}$ related  to compact binaries show up at 1PN level, whereas, as explained in Ref. \cite{Blanchet2006b}, in GR they are either  of order $\OC{-3}$ ($\mathcal{F}_{\rm SO}$) or  $\OC{-4}$ ($\mathcal{F}_{\rm SS}$ and $ \mathcal{F}_{\rm SS'}$).  This is due to the factor $\mathcal{N}$, for which we have provided a first assessment in Eq. \eqref{szi-components-NS} (see also Eq. \eqref{szi-components-BH-NS}, below).  In fact, the form of $\mathcal{N}$ can change depending on the chosen matter model  and this can influence the formal PN structure of $s_{\rm Az}$.  For a more detailed discussion,  see Sec. \ref{Sec:remarks}.

The spin-spin corrections proportional to  $s_A^2$ appear also in the angular momentum flux \eqref{ang-mom-flux-1PN} and,  to the best of our knowledge, are  not presented in the literature. We find that they are given by 
\begin{align}
\boldsymbol{\mathcal{G}}_{\rm SS'}&= \frac{8}{5}\frac{ G^3 \mu^2 M^2}{ c^7 R^6} \sum_A \biggr{[} 2 \boldsymbol{L}_{\rm N}\left(\boldsymbol{\chi}_A \cdot \boldsymbol{\chi}_A \right) - \boldsymbol{\chi}_A (\boldsymbol{\chi}_A \cdot \boldsymbol{L}_{\rm N} )  \biggr{]}.
\end{align}
The remaining contributions to $\boldsymbol{\mathcal{G}}$ being formally the same as in GR (modulo the multiplicative factor in the spin) can be read from  Ref. \cite{Kidder1995}. Similarly, the gravitational waveform \eqref{gravitational_wave_amplitude} agrees formally with that of GR and hence will not be written here explicitly (see Ref. \cite{Kidder1995}, for details). 

\subsubsection{Digression on the formal analogy between general relativity and Einstein-Cartan theory}\label{Sec:remarks}
 
In this section, we clarify some fundamental aspects of  the formal analogy  between EC and GR frameworks.

First of all, this  parallelism is formally valid at 1PN order if we consider well-separated fluid bodies having weak self-gravity, as it is confirmed by Eq. \eqref{spin-and-hat-spin}, where both $\boldsymbol{s}_A$ and  $\hat{\boldsymbol{s}}_A$ are $\OC{0}$ quantities. In this setup, in fact, both the spin-orbit and spin-spin contributions arise in GR at 1PN level both in the dynamics and the radiation field (see Refs. \cite{Kidder1995,Poisson-Will2014}, for more details). 

On the other hand, when strongly self-gravitating and maximally rotating \emph{compact objects} are taken into account, the usual procedure  exploited in GR   consists in using a variable $\hat{\boldsymbol{s}}_A$  of Newtonian order having the dimensions of an angular momentum multiplied by $c$  (see Eq. (1.1) in Ref. \cite{Blanchet2006a} and Eq. \eqref{spin-and-hat-spin-2}).  Therefore,  the leading spin-orbit and the spin-spin GR couplings pertaining to compact bodies are of order $\OC{-3}$ and  $\OC{-4}$, respectively. However, due to the different nature of the vectors $\boldsymbol{s}_A$ and $\hat{\boldsymbol{s}}_A$, we cannot  stick to the conventions employed in GR, where  $\hat{\boldsymbol{s}}_A$ has a precise and fixed form, while $\boldsymbol{s}_A$ can be described by a plethora of models.  For this reason, we have  proposed in EC theory a first  estimate for the spin vector in  Eq.  \eqref{szi-components-NS} (see also Eq. \eqref{szi-components-BH-NS}, below), where  $\boldsymbol{s}_A$ is naively of formal 3PN order. In this way, the involved PN orders are shifted, as in EC framework the leading spin-orbit and spin-spin corrections show up formally at 4PN and 7PN level, respectively. Nevertheless, the formal correspondence with GR is still recovered by means of  Eq. \eqref{spin-and-hat-spin-2}. 

However, some remarks are necessary to better explain these points. First of all, the spin vector  can be calculated  in general starting from Eq. \eqref{spin-vector-def}, where we recall $s_{ij} = \OC{0}$ \cite{Paper1,Paper2}. This means that the functional form of $\boldsymbol{s}_A$, along with its ensuing PN structure, depends on the adopted matter model. In other words, the way the integral \eqref{spin-vector-def} is performed is  influenced by the form of $s_{ij}$. 

For example, the scheme we have conceived in Eq. \eqref{szi-components-NS} (and  Eq. \eqref{szi-components-BH-NS} below) is derived  assuming that $s_{ij}$ is constant throughout the body. This brings into play the volume of the compact object, which involves either the event horizon in the black hole (BH) case or some gravitational radii for NSs. We stress that this is a rough calculation, because in more realistic models we should divide the volume of the body in regions of different densities and allow for a nonconstant $s_{ij}$.  Moreover, we should not forget the presence of the constant $\mathcal{N}$ in Eq. \eqref{szi-components-NS}, which is a novel feature with respect to  GR. This quantity indeed can be seen as a sort of compensation variable for counterbalancing the factor $c^{-6}$ occurring in Eq. \eqref{szi-components-NS}. For this reason, we have regarded both spin-orbit and spin-spin corrections as  1PN effects in EC framework also in the case of compact binaries.

This last point allows us  to discuss  another fundamental facet of our studies. In the context of compact objects,  even if we consider a new model for the spin vector $\boldsymbol{s}_A$  having a different PN structure with respect to $\hat{\boldsymbol{s}}_A$, \emph{the formal analogy between GR and EC theory still holds}, as we just need to consider Eq. \eqref{spin-and-hat-spin-2}, which shows that two frameworks are formally equivalent up to a multiplicative factor and some powers of $c$ in the spin. 

\section{Applications}
\label{sec:applications}

In this section, we apply our findings to two astrophysical situations. In Sec. \ref{sec:app_EC}, we calculate the quantum spin corrections to the energy flux and the waveform of both binary NS and BH systems. In Sec. \ref{sec:app_GR}, we propose a method to estimate the unknown macroscopic angular momentum of one of the bodies hosted in a binary system which exploits the measurement of some observed quantities and the analytical expression of the time-averaged energy flux. The main motivation behind  this last application relies  on showing that it is possible to share methodologies and results between the formally equivalent GR and EC theories. Furthermore,  we will see that the  role fulfilled by the spin-spin term \eqref{Flux-new} occurring in the energy flux is crucial. 

Like before, the spins and the angular momenta are supposed to be aligned perpendicular to the orbital plane. 

\subsection{ Quantum spin contributions to the energy flux and the gravitational waveform}
\label{sec:app_EC}

In Ref. \cite{Paper2}, we provided a first  estimate of the spin contributions to the energy flux and the gravitational waveform of both binary NSs and BHs. Our analysis was not complete, because the 1PN dynamics in EC theory was not at our disposal at that time. Therefore, we decided to set up a \emph{hybrid approach}, where the bodies were supposed to follow a GR motion parametrized by the standard Damour-Deruelle solution. Now, thanks to the results of this paper, we have all the ingredients for calculating the correct order of magnitude of the spin corrections to the gravitational  signal. As already noted in Ref. \cite{Paper2}, the new contributions to the 1PN-accurate formulas of $\mathcal{F}$ and $\mathscr{H}^{\rm TT}_{ij}$ will come only from the time derivatives of the radiative mass quadrupole moment $I_{ij}^{\rm rad}$, as the derivatives of the other moments are unchanged. 

In this treatment, we neglect any GW back-reaction effect on the source dynamics, since this hypothesis is not too restrictive as it applies to some known astrophysical GW sources (see e.g., Refs. \cite{Noutsos2020,Weisberg2010}). Moreover,  the quantum spin has the following expression  (cf. Eq.  \eqref{szi-components-NS}) \cite{Paper1,Paper2}
\begin{align}
\label{szi-components-BH-NS}
s_{Az}=\begin{cases}  \mathcal{N}\hbar\frac{4\pi}{3}\left(\frac{6Gm_A}{c^2}\right)^3, &{\rm for\ NSs},\\
\mathcal{N}\hbar\frac{4\pi}{3}\left(\frac{2Gm_A}{c^2}\right)^3, &{\rm for\ BHs},
\end{cases}.
\end{align}
In order to fulfill our goal, we define
\begin{align} \label{eq_E_F-and-E_H}
\mathcal{E}_{\mathcal{F}}(t)&:= \left|\frac{\mathcal{F}_{\rm EC}(t)}{\mathcal{F}_{\rm GR}(t)}\right|,\qquad \mathcal{E}_{\mathscr{H}} (t):=\left|\frac{\mathscr{H}_{11}^{\rm EC}(t)}{\mathscr{H}_{11}^{\rm GR}(t)}\right|.
\end{align}
Here, $\mathcal{F}_{\rm GR}:=\mathcal{F}_{\rm N}+c^{-2}\mathcal{F}_{\rm 1PN}$ and  $\mathcal{F}_{\rm EC}:=c^{-2}\left(\mathcal{F}_{\rm SO}+\mathcal{F}_{\rm SS}+\mathcal{F}_{\rm SS'}\right)$ (cf. Eq. \eqref{flux_luminosity}); similarly, we have defined $\mathscr{H}^{\rm TT}_{11}:= \mathscr{H}^{\rm GR}_{11} + \mathscr{H}^{\rm EC}_{11}$, where  $\mathscr{H}^{\rm GR}_{11}$ contains  the GR contribution, while $\mathscr{H}^{\rm EC}_{11}$ involves only the EC  terms. 

In the  hybrid scheme of Ref. \cite{Paper2}, we found for binary NSs ($2.2M_{\odot}\le M\le 4.32M_{\odot}$) $\mathcal{E}_{\mathcal{F}}\sim\mathcal{E}_{\mathscr{H}}\sim 10^{-23}$, whereas for binary BHs ($6 M_{\odot}\lesssim M \lesssim 10^{10} M_{\odot}$) we got $\mathcal{E}_{\mathcal{F}}\sim\mathcal{E}_{\mathscr{H}}\sim 10^{-13}-10^{-23}$. Now, exploiting the appropriate dynamics \eqref{eq:radius} and  the formula \eqref{eq:anal_sol} of the coordinate time  to speed up the calculations, we obtain for binary NSs $\mathcal{E}_{\mathcal{F}}\sim\mathcal{E}_{\mathscr{H}}\sim 10^{-21}$, while for binary BHs $\mathcal{E}_{\mathcal{F}}\sim\mathcal{E}_{\mathscr{H}}\sim 10^{-11}-10^{-21}$. \emph{Therefore, in the full description, the spin contributions are two orders of magnitude larger than those obtained with the hybrid approach}. This means that we can confirm the validity of our former results. In particular, in agreement with the predictions of EC model \cite{Paper2}, spin effects become more prominent when the companions get closer, as the gravitational field intensity increases. However, also with this new estimate, the EC spin corrections featuring the inspiral stage can hardly be observed with the actual and near-future GW devices. Despite that, it should be noticed that our framework relies on the simple configuration \eqref{szi-components-BH-NS}, while more sophisticated models, like those addressing the dense matter equation of state of  NSs, might be detectable.  

\subsection{Determining the macroscopic angular momentum of a companion star in a binary system}
\label{sec:app_GR}

In astrophysics, there exist binary systems composed of a primary compact object (e.g., a pulsar) and a companion star (e.g., a white  dwarf), where we know the macroscopic angular momentum $\hat{\boldsymbol{s}}_1$ of the former but not of the latter, which we denote with  $\hat{\boldsymbol{s}}_2$ (see e.g., Refs. \cite{Ridolfi2019,Corongiu2023}, for some  examples). Therefore, we propose a strategy to determine $\hat{s}_{2z}$ which exploits our analytical developments along with the measurement of the following observables: the masses of the bodies, their orbital separation $a$, the orbital period $P_{\rm b}$ and its modulation in time $\dot{P}_{\rm b}$, the Newtonian eccentricity $e_0$, and the rotation frequency $f_1$ of the primary body.  We recall that $\hat{s}_A$ can be calculated via the moment of inertia $\mathcal{I}_A$ and the  angular velocity $2\pi f_A$ of the body $A$ as  $\hat{s}_A=\mathcal{I}_A2\pi f_A$. We will suppose that  the object $A$ is a sphere of radius $\mathcal{R}_A$, so that  $\mathcal{I}_A=\frac{2}{5}m_A\mathcal{R}_A^2$. 

The crucial point of our method is that $\dot{P}_{\rm b}/P_{\rm b}$ satisfies the following relation (see Eq. (4.23) in Ref. \cite{Blanchet-Schafer1989}, for a comparison)
\begin{equation} \label{eq:Pdot_over_P}
\frac{\dot{P}_{\rm b}}{P_{\rm b}}=\frac{3}{2\mu E}\left[1-\frac{(\nu-15)}{6}\frac{E}{c^2} \right]\langle \mathcal{F}\rangle+ \OC{-10},
\end{equation}
where $\langle \mathcal{F}\rangle=\frac{1}{P_{\rm b}}\int_0^{P_{\rm b}}\mathcal{F}(t)\dd t$ is the time average of the flux. In the above equation, $E$ and $\mathcal{F}$ can be obtained directly from Eqs. \eqref{conserved-energy} and \eqref{flux_luminosity},  provided that we take into account the relation \eqref{spin-and-hat-spin-2} valid for maximally rotating compact objects.

To show how our strategy works, we consider an explicit example represented by the massive pulsar PSR J0348+0432, which is  hosted in a relativistic compact binary, where the companion star has been discovered to be a white dwarf \cite{Antoniadis2013}. This astrophysical system is gaining a lot of attention since it permits to test  gravity in the strong-field regime and allows to evaluate the orbital decay due to the GW emission. The timing parameters of PSR J0348+0432 are estimated with $1\sigma$ uncertainty by \texttt{TEMPO2} and  
are listed in Table \ref{tab:table1}.
\begin{table}[h!]
    \centering
    \begin{tabular}{|c|c|c|}
         \hline
         & &\\
        \hspace{0.3cm}{\bf PARAMETERS}\hspace{0.3cm} &\hspace{0.3cm} {\bf UNITS}\hspace{0.3cm} &\hspace{0.3cm} {\bf VALUES }\hspace{0.3cm}\\
        & &\\
         \hline \hline
        $m_1 $ & $M_\odot$ & 2.01\\
        $m_2$ & $M_\odot$ & 0.17 \\
        $\mathcal{R}_1$ & km & 17.92\\
        $\mathcal{R}_2$ & km & 45.61\\
        $f_1$ & Hz & 25.56\\
        $\hat{s}_{ 1z}$ & $10^{40}$\ J\ s & 8.31\\
        $a$ & $10^6$ km & 0.83 \\
        $e_0$ & & $2.01\times10^{-6}$ \\
        $P_{\rm b}$ & d & 0.10 \\
        $\dot{P}_{\rm b}$ & $10^{-12}$ s s${}^{-1}$ & -0.27 \\    
        \hline
        $\hat{s}_{ 2z}$ & $10^{38}$ J\ s & 2.25\\
        $f_2$ &  Hz & 0.13\\
        \hline
    \end{tabular}
    \caption{List of input parameters (taken from Ref. \cite{Antoniadis2013}) and output values  (last two rows)  of the binary system formed by the pulsar PSR J0348+0432 (labelled as body 1) and the companion white dwarf (labelled  as body 2).}
    \label{tab:table1}
\end{table}
Plugging the input data in Eq. \eqref{eq:Pdot_over_P}, we obtain a quadratic equation in $\hat{s}_{2z}$, which admits two real solutions  having opposite  sign. In our hypotheses, we choose the positive root and hence we get $\hat{s}_{2z}=2.25\times 10^{38}\ {\rm J\ s}$, which corresponds to the rotation frequency  $f_2=0.13 \ {\rm Hz}$. 

At this stage, some comments are in order. First of all, we note that the quadratic character of the  equation for $\hat{s}_{2z}$ is due to the novel  spin-spin correction \eqref{Flux-new}. Moreover, as pointed out before,   the GR contributions to the flux due to the macroscopic angular momentum occur either at 1.5PN or 2PN order when  compact objects  are investigated \cite{Blanchet2006a}. Therefore,   our treatment should involve in general the corrections appearing beyond the 1PN level which do not depend on $\hat{\boldsymbol{s}}_A$. However, these  do not alter significantly our estimate of $\hat{s}_{2z}$, and, in addition, the proposed method is still valid  since  the missing PN terms  do not change the nature of the algebraic equation to be solved to get $\hat{s}_{2z}$.

\section{Conclusions}
\label{sec:end}

Among the many proposed generalizations of GR, EC model  deals with the microphysical quantum realm and naturally fits the gauge paradigm. In this context, we have worked out the point-particle limit of the Weyssenhoff fluid and we have discovered that, at 1PN level, both the dynamical equations and the radiative multipole moments of weakly self-gravitating binary systems formally agree with the corresponding formulas framed in GR, if  the substitution \eqref{spin-and-hat-spin} is applied. The formal equivalence between GR and EC frameworks is not spoiled even if we consider compact objects, as it can be recovered via Eq.  \eqref{spin-and-hat-spin-2}. These results are not trivial if we take into account the distinct nature of GR and EC models and the different structure of the equations employed to derive them. Consider for example the case of the dynamics. The equations of motion of the Weyssenhoff fluid  stem from the \emph{generalized} conservation laws \eqref{generalized-conservation-laws}, which  lead to a \emph{generalized} Euler equation and precession motion.  These involve novel terms depending on torsion not present in GR.  Despite that, we have found that  the effacing principle holds also in EC theory and the peculiar spin-spin contact interaction of gravitational origin does not give contribution, at least at 1PN order.

Driven by the formal 1PN analogy between GR and EC frameworks, in this paper we have derived some 1PN-accurate analytical formulas pertaining to the dynamics and the radiative phenomena of  binary systems. In Sec. \ref{sec:analytical-formulae}, we have exploited the Damour-Deruelle approach to work out the relative motion of two companions having the spins and the orbital angular momentum aligned. Then, we have proposed a method to determine the function $t(\theta)$ of the coordinate time in terms of the polar angle $\theta$, which relies on the introduction of a suitably defined accumulation function. This result is crucial to speed up the evaluation of those quantities requiring the knowledge of the dynamical aspects of a binary system, such as the energy and the angular momentum fluxes, which have been displayed in Sec. \ref{Sec:Analytic-energy-and-ang-mom-fluxes}. Here, we have also shown the spin-spin couplings  \eqref{Flux-new} proportional to $s_A^2$, which, to the best of our knowledge, are not reported in the literature. 

Two applications of our analytical treatment have been given in Sec. \ref{sec:applications}. In the first one, we have improved the treatment of Ref. \cite{Paper2} and we have calculated the quantum spin contributions to the waveform and the energy flux of a binary BH and NS system. We have found that the obtained EC spin effects featuring the inspiral stage are too  weak to be observed with the actual and near-future GW apparatuses. Despite that, potentially detectable  results can be achieved during the merger stage, where the gravitational interaction is more prominent. In the second framework, we have proposed a strategy to infer the angular momentum of a body which can be useful in several astrophysical settings where it is not possible to measure the rotation frequency of both objects comprising a binary system. Our scheme requires the value of some observables, such as the orbital period modulation, the masses of the two companions, their separation, and the orbital eccentricity. Plugging these parameters in Eq. \eqref{eq:Pdot_over_P}, the unknown angular momentum is obtained by solving an algebraic second-degree equation. 

The importance of our findings consists in the fact that they can be exploited both in GR and EC frameworks thanks to their formal resemblance. In general, having analytical relations and  a mathematical methodology to obtain them can be extremely advantageous in  astrophysical contexts. In this viewpoint, our formula of $t(\theta)$ is useful for two main reasons: ($i$) it can be employed for fitting the observational data, without resorting to numerical routines, as well as gathering accurate results in acceptable times; ($ii$) the adopted strategy can be extended to higher PN orders, once a Damour-Deruelle-like solution is found. 

Our programmatic research activity has revealed some features of EC theory which, for the time being, cannot be detected. However, there are some points of our approach which need to be  improved and this  strongly spurs us to continue  our inquiry. Indeed, GWs configure to be  one of the most promising  tool to inquire new physics and  test possible deviations from GR in favour of modified frameworks \cite{Capozziello2011,Clifton2012,Bahamonde2015}.  In addition, the examination of theoretical models more adapt to describe the spin distribution inside compact objects should be developed. Beside these practical aspects, the related PN calculations can unearth new and unexpected theoretical results, apart from those highlighted in this paper.

This article can open up some interesting future perspectives. First of all, the 1PN relationship between GR and EC model and its connection to the matter field modelling the spin effects should be further explored. In fact, it should be understood  whether the link between the two theories still holds if a different fluid is considered. Moreover, the results obtained in this manuscript can be exploited to analyze radiation-reaction forces affecting the evolution of binary systems framed in EC framework. Lastly, it would be interesting to extend our approaches and methodologies also to cosmology, since it represents the natural arena for EC pattern as witnessed by the recent literature (see e.g., Refs. \cite{Medina2018,Elizalde2022,Luz2023}). These topics deserve a careful investigation in  separate papers.

\acknowledgements
Tha authors are grateful to Gruppo Nazionale di Fisica Matematica of Istituto Nazionale di Alta Matematica for support. The authors ackowkledge Prof. L. E. Kidder for the valuable correspondence. The authors thank Dr. Alessandro Ridolfi for the useful discussions on applications of our model to binary systems. The authors are grateful to the anonymous referee for having raised important conceptual issues about our work. VDF acknowledges the support of INFN {\it sez. di Napoli}, {\it iniziative specifiche} TEONGRAV. E.B. acknowledges the support of the Austrian Science Fund (FWF) grant P32086.

\bibliography{references}

\end{document}